%Paper: gr-qc/9507005
%From: ESPOSITO@napoli.infn.it
%Date: Tue, 4 Jul 1995 8:39:58 +0200 (CET-DST)

\magnification \magstep1
\raggedbottom
\openup 4\jot
\voffset6truemm
\headline={\ifnum\pageno=1\hfill\else
\hfill {\it Spectral Boundary Conditions in One-Loop ...}
\hfill \fi}
\rightline {DAMTP R-91/4}
\centerline {\bf SPECTRAL BOUNDARY CONDITIONS IN ONE-LOOP}
\centerline {\bf QUANTUM COSMOLOGY}
\vskip 1cm
\centerline {\bf Peter D. D'Eath$^{(a)}$ and Giampiero
Esposito$^{(a,b)*}$}
\vskip 1cm
\centerline {$^{(a)}$Department of Applied Mathematics and Theoretical
Physics}
\centerline {Silver Street, Cambridge CB3 9EW, U. K.}
\centerline {$^{(b)}$St. John's College, Cambridge CB2 1TP, U. K.}
\vskip 1cm
\centerline {\bf Abstract}
\vskip 1cm
\noindent
For fermionic fields on a compact Riemannian manifold with boundary one has
a choice between local and non-local (spectral) boundary conditions. The
one-loop prefactor in the Hartle-Hawking amplitude in quantum cosmology
can then be studied using the generalized Riemann $\zeta$-function formed
from the squared eigenvalues of the four-dimensional fermionic operators.
For a massless Majorana spin-${1\over 2}$
field, the spectral conditions involve setting to zero half of the
fermionic field on the boundary, corresponding to harmonics of the intrinsic
three-dimensional Dirac operator on the boundary with positive eigenvalues.
Remarkably, a detailed calculation for the case of a flat background
bounded by a three-sphere yields the same value $\zeta(0)={11\over 360}$ as
was found previously by the authors
using local boundary conditions. A similar calculation for a
spin-${3\over 2}$ field, working only with physical degrees of freedom
(and, hence, excluding gauge and ghost modes, which contribute to the full
Becchi-Rouet-Stora-Tyutin-invariant
amplitude), again gives a value $\zeta(0)=-{289\over 360}$
equal to that for the natural local boundary conditions.
\vskip 14cm
\leftline {Phys. Rev. D {\bf 44}, 1713 (1991).}
\vskip 25cm
\centerline {\bf I. INTRODUCTION}
\vskip 1cm
In recent work by the authors,$^{1,2}$ one-loop effects in quantum cosmology
for fermionic fields have been studied using local boundary conditions. In
the case of a massless spin-${1\over 2}$ field
$\Bigr(\psi^{A}, \; {\widetilde \psi}^{A'}\Bigr)$ in a Riemannian
background, which we shall refer to loosely as a Majorana spin-${1\over 2}$
field,$^{2}$ the simplest natural local boundary conditions are
(using two-component spinors)
$$
\sqrt{2} \; {_{e}n_{A}^{\; \; A'}}\psi^{A}=\epsilon \; {\widetilde \psi}^{A'}
\; \; \; \; ,
\eqno (1.1)
$$
where ${_{e}n^{AA'}}$ is the spinor version of the unit Euclidean normal
${_{e}n^{\mu}}$ to the boundary and $\epsilon=\pm 1$. Note that the primed
field ${\widetilde \psi}^{A'}$ is taken to be independent of $\psi^{A}$, not
related by any conjugation operation. A first-order differential operator
for this Riemannian boundary-value problem exists which is symmetric and
has self-adjoint extensions. One can then study the generalized Riemann
zeta-function formed from the squared eigenvalues of the Dirac operator.
The value of $\zeta(0)$ yields the one-loop divergence of the quantum
amplitude for the Hartle-Hawking quantum state subject to these boundary
conditions. Further, $\zeta(0)$ determines the scaling of the one-loop
amplitude : in the case of a flat Euclidean four-dimensional background
geometry bounded by a three-sphere of radius $a$, the one-loop amplitude
scales as $a^{-\zeta(0)}$ for a fermionic field (in the case of a
scale-independent measure). A direct calculation$^{2}$ for a massless
Majorana spin-${1\over 2}$ field with the boundary conditions (1.1) on a
three-sphere in flat four-space gave the value $\zeta(0)={11\over 360}$.
Such boundary conditions are of interest because they are part of a
supersymmetric family of local boundary conditions for both fermions and
bosons,$^{1-5}$ so that one can check whether or not the one-loop divergences
in the Hartle-Hawking amplitude cancel in extended supergravity theories.

Because of the first-order nature of the fermionic operators, one has a
choice between local boundary conditions such as Eq. (1.1) and
non-local (spectral) boundary conditions. While spectral boundary conditions
are not in any obvious way related to supersymmetry, they are nevertheless
of considerable mathematical interest, and are the subject of this paper.
Their mathematical foundations lie in the theory of elliptic equations
and in the index theory for the Dirac operator.$^{6}$ To illustrate these
boundary conditions, consider again the case of a massless Majorana
spin-${1\over 2}$ field $\Bigr(\psi^{A}, \; {\widetilde \psi}^{A'}\Bigr)$ in
the region of flat Euclidean four-space bounded by a three-sphere of radius
$a$. Denote by $\tau$ the Euclidean distance from the centre of the sphere.
Then the field $\Bigr(\psi^{A}, \; {\widetilde \psi}^{A'}\Bigr)$ may be
expanded in terms of harmonics on the family of spheres centred on the
origin,$^{2,7}$ as
$$
\psi^{A}={\tau^{-{3\over 2}}\over 2\pi}\sum_{npq}\alpha_{n}^{pq}
\Bigr[m_{np}(\tau)\rho^{nqA}+{\widetilde r}_{np}(\tau)
{\overline \sigma}^{nqA}\Bigr] \; \; \; \; ,
\eqno (1.2)
$$
$$
{\widetilde \psi}^{A'}={\tau^{-{3\over 2}}\over 2\pi}\sum_{npq}\alpha_{n}^{pq}
\left[{\widetilde m}_{np}(\tau){\overline \rho}^{nqA'}+
r_{np}(\tau)\sigma^{nqA'}\right] \; \; \; \; .
\eqno (1.3)
$$
In the summations, $n$ runs from $0$ to $\infty$, $p$ and $q$ from $1$ to
$(n+1)(n+2)$. The $\alpha_{n}^{pq}$ are a collection of matrices introduced
for convenience, where, for each $n$, $\alpha_{n}^{pq}$ is block-diagonal
in the indices $pq$, with blocks
$\pmatrix { 1&1\cr 1&-1\cr}$. The harmonics $\rho^{nqA}$ have positive
eigenvalues ${1\over 2}(n+{3\over 2})$ of the intrinsic three-dimensional
Dirac operator ${_{e}n_{AA'}}e^{BA'j}{ }^{(3)}D_{j}$ on the three-sphere,
while the harmonics ${\overline \sigma}^{nqA}$ have negative eigenvalues
$-{1\over 2}(n+{3\over 2})$. Here $e^{BA'j}$ is the spinor version of the
orthonormal spatial triad on the three-sphere, and ${ }^{(3)}D_{j}$ is the
three-dimensional covariant derivative $(j=1,2,3)$.$^{7}$ Similarly, the
harmonics $\sigma^{nqA'}$ have positive eigenvalues
${1\over 2}(n+{3\over 2})$ of the corresponding three-dimensional operator
on primed spinors, and the harmonics ${\overline \rho}^{nqA'}$ have
negative eigenvalues $-{1\over 2}(n+{3\over 2})$. This expansion can be
summarized more simply as
$$
\psi^{A}=\psi_{(+)}^{A}+\psi_{(-)}^{A} \; \; \; \; ,
\eqno (1.4)
$$
$$
{\widetilde \psi}^{A'}={\widetilde \psi}_{(-)}^{A'}+
{\widetilde \psi}_{(+)}^{A'} \; \; \; \; ,
\eqno (1.5)
$$
where the $(\pm)$ parts correspond to positive and negative eigenvalues,
respectively, for the intrinsic three-dimensional Dirac operator.

In studying the classical boundary-value problem for the massless Dirac
equation, one finds that classical solutions corresponding to boundary
data with a non-zero coefficient ${\widetilde r}_{np}(a)$ or
${\widetilde m}_{np}(a)$ diverge as a negative power of $\tau$ near the
origin.$^{7}$ Boundary data with a non-zero coefficient of $m_{np}(a)$ or
$r_{np}(a)$ yield a regular solution of the massless Dirac equation,
proportional to a positive power of $\tau$. Thus the classical
boundary-value problem is only well-posed if one specifies the $m_{np}(a)$
and $r_{np}(a)$, but not the remaining data.
In the case of a general manifold with boundary,
knowledge of the spectrum of the intrinsic three-dimensional Dirac
operator is necessary if one wishes to compute the $\eta$-invariant which
gives a boundary contribution to the index of the Dirac operator for the
manifold with boundary.$^{6}$ [In the generic case, the index
will be non-zero, and the classical boundary-value problem will
not be well-posed.$^{6}$] In this paper we are however not concerned
with the signature, but rather, as explained above, with the
zeta-function $\zeta(s)$ formed from the eigenvalues of the
four-dimensional Dirac operator, subject now to boundary conditions in
which $\psi_{(+)}^{A}$ and ${\widetilde \psi}_{(+)}^{A'}$ are specified
on the boundary in our flat-space example.

Thus, just as one has a well-posed classical problem with these boundary
data, one similarly expects that the analogous quantum amplitude, the
Hartle-Hawking path integral
$$
K_{HH}=\int e^{-I_{E}}D\psi^{A} \; D{\widetilde \psi}^{A'}
\eqno (1.6)
$$
for the fermions, is naturally studied by taking spectral boundary
conditions in which $\psi_{(+)}^{A}$ and ${\widetilde \psi}_{(+)}^{A'}$,
or equivalently the $m_{np}(a)$ and $r_{np}(a)$ in our example, are
specified on the boundary. Here the Euclidean action is
$$
I_{E}={i\over 2}\int d^{4}x \; \sqrt{g}
\left[{\widetilde \psi}^{A'}\left(\nabla_{AA'}\psi^{A}\right)
-\left(\nabla_{AA'}{\widetilde \psi}^{A'}\right)\psi^{A}\right]
+I_{B} \; \; \; \; .
\eqno (1.7)
$$
The fermionic fields are taken to be anti-commuting, and Berezin integration
is being used.$^{7}$ With our conventions, the Infeld-van der Waerden
connection symbols $\sigma_{a}^{AA'}$ are taken to be
$\sigma_{0}=-{i\over \sqrt{2}}I$, $\sigma_{i}={\Sigma_{i}\over \sqrt{2}}$
($i=1,2,3$), where $\Sigma_{i}$ are the Pauli matrices. A boundary term
$I_{B}$, discussed in Ref. 7, is needed in general. In our simple example
of the three-sphere, the Hartle-Hawking amplitude is then a function
$K_{HH}\Bigr[m_{np}(a),r_{np}(a)\Bigr]$ of the spectral boundary data.
The one-loop properties of this amplitude can be studied without loss of
generality by setting the allowed boundary data $m_{np}(a)$, $r_{np}(a)$
to zero, $\forall n,p$, so that the boundary conditions become
$$
\psi_{(+)}^{A}=0 \; \; \; \; , \; \; \; \;
{\widetilde \psi}_{(+)}^{A'}=0 \; \; \; \; .
\eqno (1.8)
$$
[A similar simplification was made in Ref. 2 in choosing the local
boundary conditions (1.1), rather than specifying a non-zero spinor field
$\sqrt{2}\;{_{e}n_{A}^{\; \; A'}}\psi^{A}-\epsilon \; {\widetilde \psi}^{A'}$
on the boundary.] The boundary term $I_{B}$ is zero in this case.

In Sec. II the action (1.7) is expanded in harmonics, subject to the
spectral boundary conditions (1.8) on the three-sphere of radius $a$
bounding a region of flat four-space. The eigenvalue equation arising in
the evaluation of the one-loop functional determinant is derived. The
resulting zeta-function formed from the squared eigenvalues is related
to the heat kernel $G(T)$ for the Laplacian operator on spinors. In Sec. III
the Laplace transforms of the corresponding Green's functions are derived.
The detailed calculations leading to the asymptotic expansion of $G(T)$ as
$T \rightarrow 0^{+}$ are described in Sec. IV. These lead to the value
$\zeta(0)={11\over 360}$, which remarkably is the same as that found
previously$^{2}$ for a massless Majorana spin-${1\over 2}$ field
using the local boundary conditions (1.1) on the same manifold with
boundary. An analogous calculation for the spin-${3\over 2}$ field is
sketched in Sec. V, working only with the physical degrees of freedom in a
particular gauge. This of course excludes the contribution of gauge and
ghost modes which should appear in the full
Becchi-Rouet-Stora-Tyutin-invariant path integral.
Nevertheless, it is striking that the value $\zeta(0)=-{289\over 360}$
obtained is again identical to that found using the natural local boundary
conditions for the physical degrees of freedom.$^{1}$ Some comments are
included in Sec. VI.
\vskip 20cm
\centerline {\bf II. EIGENVALUES FOR SPECTRAL BOUNDARY}
\centerline {\bf CONDITIONS ON $S^3$}
\vskip 1cm
In the case of a region of four-dimensional flat Euclidean space bounded
by a three-sphere of radius $a$, we decompose the massless spin-${1\over 2}$
field $\Bigr(\psi^{A}, \; {\widetilde \psi}^{A'}\Bigr)$ as in Eqs. (1.2,3)
and impose the spectral boundary conditions (1.8), so that
$$
m_{np}(a)=0 \; \; \; \; , \; \; \; \; r_{np}(a)=0 \; \; \; \; ,
\; \; \; \; \forall n,p \; \; \; \; .
\eqno (2.1)
$$
The Hartle-Hawking path integral (1.6), with the Euclidean action $I_{E}$
given by Eq. (1.7), can then be studied, with $m_{np}(\tau),r_{np}(\tau)$
constrained by Eq. (2.1) at the boundary, but ${\widetilde m}_{np}(a)$
and ${\widetilde r}_{np}(a)$ unconstrained. The physical fields
$\Bigr(\psi^{A}, \; {\widetilde \psi}^{A'}\Bigr)$ summed over in the path
integral should at least be bounded near the origin $\tau=0$. Because of
the factor $\tau^{-{3\over 2}}$ in Eqs. (1.2,3), this implies that
$$
m_{np}(0)=r_{np}(0)={\widetilde m}_{np}(0)={\widetilde r}_{np}(0)=0
\; \; \; \; , \; \; \; \; \forall n,p \; \; \; \; .
\eqno (2.2)
$$
The action $I_{E}$ can then be expanded out in terms of harmonics, by
analogy with the treatment of Ref. 7, as
$$
I_{E}=\sum_{n=0}^{\infty}\sum_{p=1}^{(n+1)(n+2)}
\Bigr[I_{n}(m_{np},{\widetilde m}_{np})+I_{n}(r_{np},{\widetilde r}_{np})
\Bigr] \; \; \; \; ,
\eqno (2.3)
$$
where
$$
I_{n}(x,{\widetilde x})=\int_{0}^{a}d\tau \; \left[
{1\over 2}\Bigr({\widetilde x}{\dot x}+x{\dot {\widetilde x}}\Bigr)-
{(n+{3\over 2})\over \tau}{\widetilde x}x \right] \; \; \; \; ,
\eqno (2.4)
$$
and an overdot denotes ${d\over d\tau}$.
Note that, as remarked in the Introduction, no boundary term $I_{B}$ of the
type described in Ref. 7 appears in the action, because of the boundary
conditions (2.1,2).

Because of the degeneracy $(n+1)(n+2)$ in the label $p$, and because of the
splitting of the action $I_{E}$ in Eq. (2.3) into identical pieces
involving $(m_{np},{\widetilde m}_{np})$ and $(r_{np},{\widetilde r}_{np})$,
the complete path integral (1.6) splits into a product of Berezin
integrals :
$$
K_{HH}=\prod_{n=0}^{\infty}
{\Bigr \{ \int d[x]d[{\widetilde x}]exp[-I_{n}(x,{\widetilde x})]\Bigr\} }
^{2(n+1)(n+2)}
\; \; \; \; .
\eqno (2.5)
$$
The boundary conditions in each integration are then
$$
x(0)={\widetilde x}(0)=0 \; \; \; \; , \; \; \; \; x(a)=0 \; \; \; \; ,
\eqno (2.6)
$$
following Eqs. (2.1,2).

Equivalently, one can follow the procedure of Ref. 2 and study the eigenvalue
equations
$$
\nabla_{AA'}\psi_{m}^{A}=\lambda_{m}{\widetilde \psi}_{mA'} \; \; \; \; ,
\; \; \; \; \nabla_{AA'}{\widetilde \psi}_{m}^{A'}=\lambda_{m}
\psi_{mA} \; \; \; \; ,
\eqno (2.7)
$$
naturally arising from variation of the action (1.7), subject to the
boundary conditions (1.8). The eigenfunctions $\Bigr(\psi_{m}^{A}, \;
{\widetilde \psi}_{m}^{A'}\Bigr)$ are clearly found by separation of
variables, being of the form
$$
\psi_{n}^{A}=\tau^{-{3\over 2}}x_{nk}(\tau)\rho^{npA} \; \; \; \; ,
\; \; \; \; {\widetilde \psi}_{n}^{A'}=\tau^{-{3\over 2}}
{\widetilde x}_{nk}(\tau){\overline \rho}^{npA'}
\; \; \; \; ,
\eqno (2.8)
$$
or
$$
\psi_{n}^{A}=\tau^{-{3\over 2}}{\widetilde x}_{nk}(\tau)
{\overline \sigma}^{npA} \; \; \; \; , \; \; \; \;
{\widetilde \psi}_{n}^{A'}=\tau^{-{3\over 2}}x_{nk}(\tau)\sigma^{npA'}
\; \; \; \; .
\eqno (2.9)
$$
Here the pair $\Bigr(x_{nk}(\tau), \; {\widetilde x}_{nk}(\tau)\Bigr)$
is an eigenvector corresponding to variation of the action
$I_{n}(x,{\widetilde x})$ of Eqs. (2.4,5), obeying
$$
\left({d\over d\tau}-\nu\right)x_{nk}=E_{nk}{\widetilde x}_{nk}
\; \; \; \; ,
\; \; \; \; \left(-{d\over d\tau}-\nu \right){\widetilde x}_{nk}=
E_{nk}x_{nk} \; \; \; \; ,
\eqno (2.10,11)
$$
subject to the boundary conditions $x_{nk}(0)={\widetilde x}_{nk}(0)=0$,
$x_{nk}(a)=0$, where $\nu={(n+{3\over 2})\over \tau}$ and
$E_{nk}=i\lambda_{nk}$. For each $n$, the index $k$ labels the countable
set of corresponding eigen-functions and -values.

The Gaussian fermionic path integral (2.5) is then formally proportional
to the product of the eigenvalues
$\prod_{n,k}{\left({\lambda_{n,k}\over {\widetilde \mu}}\right)}
^{2(n+1)(n+2)}$. Here the constant ${\widetilde \mu}$ with dimensions of mass
has been introduced in order to make the product dimensionless.$^{2}$ As
shown below, the $\lambda_{n,k}$ are purely imaginary, or equivalently the
$E_{nk}$ are real. Further, the $E_{nk}$ occur in equal and opposite pairs,
since if the pair $\Bigr(x_{nk}(\tau), \; {\widetilde x}_{nk}(\tau)\Bigr)$
corresponds to an eigenvalue $E_{nk}$ in Eq. (2.10), then clearly the pair
$\Bigr(x_{nk}(\tau), -{\widetilde x}_{nk}(\tau)\Bigr)$ corresponds to the
eigenvalue $-E_{nk}$. Hence the path integral can also be written as
$\prod_{n,k}{\left({{\mid \lambda_{nk} \mid}\over {\widetilde \mu}}\right)}
^{2(n+1)(n+2)}$. The Berezin integration rules imply that one should only
include those values of $k$, say $k=1,2,...$, which correspond to positive
values of $E_{nk}$. This formal expression must then be regularized using
zeta-function methods.

The coupled first-order equations (2.10,11) lead to the second-order
equation
$$
\left({d^{2}\over d\tau^{2}}-{[(n+1)^{2}-{1\over 4}]\over \tau^{2}}+
E_{nk}^{2}\right)x_{nk}=0 \; \; \; \; ,
\eqno (2.12)
$$
together with a corresponding equation for ${\widetilde x}_{nk}(\tau)$. The
solutions $\Bigr(x_{nk}(\tau), \; {\widetilde x}_{nk}(\tau)\Bigr)$ obeying
the boundary conditions (2.6) are
$$
x_{nk}=A_{nk}\sqrt{\tau}J_{n+1}(E_{nk}\tau) \; \; \; \; ,
\eqno (2.13)
$$
$$
{\widetilde x}_{nk}=-A_{nk}\sqrt{\tau}J_{n+2}(E_{nk}\tau) \; \; \; \; ,
\eqno (2.14)
$$
subject to the eigenvalue condition
$$
J_{n+1}(E_{nk}a)=0 \; \; \; \; , \; \; \; \;
n=0,1,2,... .
\eqno (2.15)
$$
The quantities $E_{nk}^{2}$ are clearly real and positive since they are
eigenvalues for the self-adjoint problem (2.12), subject to boundary
conditions $x_{nk}(0)=x_{nk}(a)=0$, and since Eq. (2.12) involves a
positive operator. For a given $n$, the coefficients $A_{nk}$ in Eqs.
(2.13,14) can be chosen such that the eigenfunctions $x_{nk}(\tau)$ are
orthonormal in the inner product $(u,v)=\int_{0}^{a}d\tau \; u(\tau)
v(\tau)$, as are the corresponding ${\widetilde x}_{nk}(\tau)$. For each
$n$, the action $I_{n}(x,{\widetilde x})$ of Eq. (2.4) becomes a diagonal
sum over the eigenfunctions $x_{nk},{\widetilde x}_{nk}$ for
$k=1,2,...$. Performing the Berezin integrations in Eq. (2.5), one arrives
at the formal expression quoted in the previous paragraph :
$$
K_{HH}=\prod_{n=0}^{\infty}\prod_{k=1}^{\infty}
{\left({E_{nk}\over {\widetilde \mu}}\right)}^{2(n+1)(n+2)}
\; \; \; \; ,
\eqno (2.16)
$$
where the $E_{nk}$ ($k=1,2,...$) are the positive eigenvalues obeying Eq.
(2.15).

Following the standard procedure, as, for example, in Ref. 2, the formally
divergent infinite product (2.16) is regularized by studying the
zeta-function for the squared eigenvalues
$$
\zeta(s)=\sum_{n=0}^{\infty}\sum_{k=1}^{\infty}d_{k}(n)
(E_{nk})^{-2s} \; \; \; \; .
\eqno (2.17)
$$
Here the degeneracy $d_{k}(n)=2(n+1)(n+2)$ is in fact independent of $k$.
The series (2.17) converges for $Re(s)>2$, and can be analytically
continued to a meromorphic function with poles only at
$s={1\over 2},1,{3\over 2},2$. The formal expression $\log(K_{HH})$,
with $K_{HH}$ given by Eq. (2.16), is then evaluated as
$-{1\over 2}\zeta'(0)-\zeta(0) \log {\widetilde \mu}$.
\vskip 1cm
\centerline {\bf III. GREEN'S FUNCTIONS AND THE HEAT KERNEL}
\vskip 1cm
The quantity $\zeta(0)$, which gives the divergence and scaling properties of
the one-loop amplitude, is evaluated by studying the heat kernel, defined
for $T>0$ by
$$
G(T)=\sum_{n=0}^{\infty}\sum_{k=1}^{\infty}2(n+1)(n+2)e^{-(E_{nk})^{2}T}
\; \; \; \; .
\eqno (3.1)
$$
This is related to the zeta-function $\zeta(s)$ of Eq. (2.17) by
$$
\zeta(s)={1\over \Gamma(s)}\int_{0}^{\infty}dT \; T^{s-1}G(T) \; \; \; \;
\eqno (3.2)
$$
for $Re(s)>2$. The heat kernel $G(T)$ has the standard asymptotic expansion
$$
G(T) \sim \sum_{i=0}^{\infty}B_{i}T^{{i\over 2}-2} \; \; \; \; ,
\eqno (3.3)
$$
as $T \rightarrow 0^{+}$, where in particular $B_{4}=\zeta(0)$.$^{8}$

In the present case of spectral boundary conditions, the eigenvalue
condition
$$J_{n+1}(E_{nk}a)=0 \; \; \; \; , \; \; \; \; n=0,1,2,...$$
[Eq. (2.15)], with degeneracy
$2(n+1)(n+2)$, is considerably simpler than the eigenvalue condition
$[J_{n+1}(E_{nk}a)]^{2}-[J_{n+2}(E_{nk}a)]^{2}=0$,
$n=0,1,2,...$, with degeneracy $(n+1)(n+2)$, found in Ref. 2 for a
spin-${1\over 2}$ field with local boundary conditions on $S^3$. As a
consequence, we can use a more straightforward treatment, following
(among others) Schleich$^{8}$ and Stewartson and Waechter,$^{9}$ which
involves the Green's function for the heat equation for each $n$. The
value $\zeta(0)={11\over 360}$ which results is, perhaps surprisingly,
equal to that found in Ref. 2 for local boundary conditions.

One proceeds by considering, for each $n=0,1,2,...$, the Green's function
defined for $T>0$ by
$$
G_{n}(\tau,\tau',T)=\sum_{k=1}^{\infty}x_{nk}(\tau)x_{nk}(\tau')
e^{-(E_{nk})^{2}T} \; \; \; \; ,
\eqno (3.4)
$$
with $G_{n}(\tau,\tau',T)=0$ for $T\leq 0$. Here the $x_{nk}(\tau)$ are the
eigenfunctions of Eq. (2.12), obeying Eq. (2.12) and
$x_{nk}(0)=x_{nk}(a)=0$, and normalized according to
$\int_{0}^{a}d\tau \; x_{nk}(\tau)x_{nl}(\tau)=\delta_{kl}$, as described
in Sec. II. Here $G_{n}(\tau,\tau',T)$ is the Green's function for the
heat equation
$$
\left[{\partial \over \partial T}-{\partial^{2}\over \partial \tau^{2}}
+{((n+1)^{2}-{1\over 4})\over \tau^{2}}\right]G_{n}(\tau,\tau',T)=
\delta(\tau-\tau')\delta(T) \; \; \; \; .
\eqno (3.5)
$$
It obeys the boundary conditions
$$
G_{n}(a,\tau',T)=G_{n}(\tau,a,T)=G_{n}(0,\tau',T)=G_{n}(\tau,0,T)=0
\; \; \; \; .
\eqno (3.6)
$$

By setting $\tau=\tau'$ and integrating, one recovers the contribution
$$
G_{n}(T)=\int_{0}^{a}d\tau \; G_{n}(\tau,\tau,T)=\sum_{k=1}^{\infty}
e^{-(E_{nk})^{2}T}
\eqno (3.7)
$$
to the heat kernel
$$
G(T)=\sum_{n=0}^{\infty}2(n+1)(n+2)G_{n}(T) \; \; \; \; .
\eqno (3.8)
$$

The Laplace transform of the Green's function,
$$
{\hat G}_{n}(\tau,\tau',\sigma^{2})=\int_{0}^{\infty}dT \;
e^{-{\sigma}^{2}T}G_{n}(\tau,\tau',T) \; \; \; \; ,
\eqno (3.9)
$$
obeys the differential equation
$$
\left[{\partial^{2}\over \partial \tau^{2}}-\sigma^{2}-
{((n+1)^{2}-{1\over 4})\over \tau^{2}}\right]{\hat G}_{n}
(\tau,\tau',\sigma^{2})=-\delta(\tau-\tau') \; \; \; \; .
\eqno (3.10)
$$
Following Eq. (3.6), ${\hat G}_{n}(\tau,\tau',\sigma^{2})$ is zero
whenever either $\tau$ or $\tau'$ is $0$ or $a$. It can be found explicitly
in terms of modified Bessel functions (cf Refs. 8-11), as
$$ \eqalignno{
{\hat G}_{n}(\tau,\tau',\sigma^{2})&=
(\tau_{<})^{1\over 2}(\tau_{>})^{1\over 2}{I_{n+1}(\sigma \tau_{<})\over
I_{n+1}(\sigma a)}\cr
&\Bigr[I_{n+1}(\sigma a)K_{n+1}(\sigma \tau_{>})-
I_{n+1}(\sigma \tau_{>})K_{n+1}(\sigma a)\Bigr] \; \; \; \; ,
&(3.11)\cr}
$$
where $\tau_{>}$ ($\tau_{<}$) is the larger (smaller) of $\tau$ and
$\tau'$. This gives a splitting of the Laplace transform
${\hat G}(\tau,\tau',\sigma^{2})$ of the function $G(\tau,\tau',T)$,
defined as
$$
G(\tau,\tau',T)=\sum_{n=0}^{\infty}2(n+1)(n+2)G_{n}(\tau,\tau',T)
\; \; \; \; ,
\eqno (3.12)
$$
in the form
$$
{\hat G}(\tau,\tau',\sigma^{2})={\hat G}^{F}(\tau,\tau',\sigma^{2})
+{\hat G}^{I}(\tau,\tau',\sigma^{2}) \; \; \; \; .
\eqno (3.13)
$$
Here
$$
{\hat G}^{F}(\tau,\tau',\sigma^{2})=\sum_{n=0}^{\infty}2(n+1)(n+2)
(\tau_{<})^{1\over 2}(\tau_{>})^{1\over 2}I_{n+1}(\sigma \tau_{<})
K_{n+1}(\sigma \tau_{>})
\eqno (3.14)
$$
is the ``free contribution'',
which corresponds to the boundary conditions of vanishing at
the origin and at infinity. The ``interacting contribution'' is
$$
{\hat G}^{I}(\tau,\tau',\sigma^{2})=-\sum_{n=0}^{\infty}2(n+1)(n+2)
(\tau_{<})^{1\over 2}(\tau_{>})^{1\over 2}{K_{n+1}(\sigma a)\over
I_{n+1}(\sigma a)}I_{n+1}(\sigma \tau_{<})I_{n+1}(\sigma \tau_{>}) \; .
\eqno (3.15)
$$

By studying the large-$\sigma^{2}$ behaviour of these functions, or the
corresponding small-$T$ behaviour of $G(T)$ as in Eq. (3.3), one finds
$\zeta(0)$.
\vskip 1cm
\centerline {\bf IV. DETAILED CALCULATION OF THE INFINITE SUMS FOR
SPIN ${1\over 2}$}
\vskip 1cm
Following the calculation of Schleich,$^{8}$ taking the inverse Laplace
transform term by term and integrating with respect to $\tau$ as in Eqs.
(3.7,8), the free part of the heat kernel is found to be
$$
G^{F}(T)= \int_{0}^{{\textstyle {a^2}}\over 2T}
\sum_{n=1}^{\infty}\Bigr[n(n+1)\Bigr]
I_{n}(y) e^{-y} \; dy
\; \; \; \; .
\eqno (4.1)
$$
Using the integral representation of the Bessel functions,$^{11}$ one obtains
the identity
$$
\sum_{n=1}^{\infty}n^{2}I_{n}(y)=
{y\over 2}e^{y} \; \; \; \; ,
\eqno (4.2)
$$
which implies
$$
\sum_{n=1}^{\infty}\int_{0}^{{\textstyle {a^2}}\over 2T}
n^{2}I_{n}(y)e^{-y}\; dy =
{a^4 \over 16 T^2} \; \; \; \; .
\eqno (4.3)
$$
Moreover, using the relations$^{11}$
$$
I_{0}'(y)=I_{1}(y) \; \; \; \; , \; \; \; \;
nI_{n}(y)={y\over 2}\Bigr(I_{n-1}(y)-I_{n+1}(y)\Bigr)
\eqno (4.4,5)
$$
one has
$$ \eqalignno{
\int_{0}^{{\textstyle {a^2}}\over 2T}
\sum_{n=1}^{\infty}nI_{n}(y)e^{-y} \; dy &=
\int_{0}^{{\textstyle {a^2}}\over 2T}
\sum_{n=1}^{\infty}\Bigr(I_{n-1}(y)-I_{n+1}(y)\Bigr){y\over 2}e^{-y} \; dy \cr
&={1\over 2}\int_{0}^{{\textstyle {a^2}}\over 2T}
ye^{-y}I_{0}(y)\; dy + {1\over 2}\int_{0}^{{\textstyle {a^2}}\over 2T}
ye^{-y}{dI_{0}\over dy}\; dy \cr
&={1\over 2}{a^2 \over 2T} e^{-{{\textstyle {a^2}}\over 2T}}
I_{0}\left({a^2 \over 2T}\right)
\cr &+ {1\over 2}\int_{0}^{{\textstyle {a^2}}\over 2T}
e^{-y}(2y-1)I_{0}(y)dy \; \; \; \; .
&(4.6) \cr}
$$
The problem of computing $G^{F}(T)$ is thus reduced to that of computing
the right-hand side of Eq. (4.6). Using again the identity
$I_{0}'(y)=I_{1}(y)$, together with $I_{1}'(y)=I_{0}(y)-y^{-1}I_{1}(y)$,
one has the relations among indefinite integrals :
$$ \eqalignno{
\int ye^{-y}I_{0}(y) \; dy &=
\int \left[ye^{-y}(I_{0}+I_{1})-y^2 e^{-y}(I_{0}+I_{1}) +y^2 e^{-y}
\left(I_{1}+I_{0}-{I_{1}\over y}\right)\right] \; dy \cr
&=y^2 e^{-y}(I_{0}+I_{1})-\int ye^{-y}(I_{0}+I_{1})\; dy
\; \; \; \; , &(4.7) \cr}
$$
giving
$$
3\int ye^{-y}I_{0}(y)\; dy =y^2 e^{-y}(I_{0}+I_{1})-ye^{-y}I_{0}+
\int I_{0}e^{-y} \; dy \; \; .
\eqno (4.8)
$$
In addition,
$$ \eqalignno{
\int I_{0}e^{-y}\; dy&=
\int {\left[ye^{-y}\left(I_{1}+I_{0}-{I_{1}\over y}\right)
+\left(I_{0}+I_{1}\right)
\left(e^{-y}-ye^{-y}\right)\right]} \; dy \cr
&=ye^{-y}(I_{0}+I_{1}) \; \; \; \; . &(4.9)\cr}
$$
Hence the integral on the right-hand side of Eq. (4.6) is found to be
$$
\int e^{-y}(2y-1)I_{0}(y)\; dy = e^{-y}\left\{{2\over 3}\Bigr[y^2 I_{0} +
(y^2 +y)I_{1}\Bigr] -y(I_{0}+I_{1})\right\} \; \; .
\eqno (4.10)
$$
The relations (4.1)-(4.6) and (4.10) imply that,
as $T \rightarrow 0^{+}$
$$
G^{F}(T) \sim {a^4 \over 16}T^{-2}+
{a^3 \over 6\sqrt{\pi}}T^{-{3\over 2}}
-{a\over 8\sqrt{\pi}}
T^{-{1\over 2}} +O(\sqrt{T}) \; \; \; \; .
\eqno (4.11)
$$
In deriving Eq. (4.11), we have used the following asymptotic relations$^{11}$
valid as ${\textstyle
{z \rightarrow \infty}}$ :
$$
e^{-z}z^2 I_{0}(z)\sim {z^{\textstyle {3\over 2}}\over \sqrt{2\pi}}
+{1\over 8}
{\sqrt{z}\over \sqrt{2\pi}} + O \left({1\over \sqrt{z}}\right) \; \; \; \; ,
\eqno (4.12)
$$
$$
e^{-z}z^{2}I_{1}(z)\sim {z^{\textstyle {3\over 2}}\over \sqrt{2\pi}}-
{3\over 8}
{\sqrt{z}\over \sqrt{2\pi}} + O \left({1\over \sqrt{z}}\right) \; \; \; \; .
\eqno (4.13)
$$

The Laplace transform of the kernel of the interacting part is given by
$$
G^{I}(\sigma^2)=-a^{2} \sum_{n=1}^{\infty}n^2 f(n;\sigma a)
-a^{2} \sum_{n=1}^{\infty}nf(n;\sigma a) \; \; \; \; ,
\eqno (4.14)
$$
where
$$
f(n;\sigma a)=\left(1+{n^{2}\over \sigma^{2}a^{2}}\right)I_{n}(\sigma a)
K_{n}(\sigma a)-I_{n}'(\sigma a)K_{n}'(\sigma a)
-{I_{n}'(\sigma a)\over \sigma a I_{n}(\sigma a)}
\eqno (4.15)
$$
is the function defined in Eq. (5.9) of Ref. 8, and in Ref. 9. In fact
(see Kennedy$^{12}$) the sums $\sum_{n=1}^{\infty}$ in Eq. (4.14) diverge
because of the factors of $n^{2}$ and $n$. This occurs because we are
attempting to take the Laplace transform of a function $G(T)$ which is
singular as $T \rightarrow 0$ [see Eq. (3.3)]. This difficulty can be
avoided by first computing the sums $\sum_{n=1}^{N}$ for large $\sigma^{2}$,
using the asymptotic expansion of $f(n;\sigma a)$ valid uniformly$^{8-10}$
with respect to $n$ at large $\sigma a$, and then taking the inverse
Laplace transform before taking the limit $N \rightarrow \infty$. The first
series in Eq. (4.14) has already been studied in the case of scalar
fields.$^{9}$ In the case of the second series, the Watson transform used
in Refs. 8-9 is a source of complications, because $nf(n;\sigma a)$ is not
an even function of $n$. Instead, we take the inverse Laplace transform
of the large-${\sigma}^{2}$ expansion of $nf(n;\sigma a)$, and compute the
sum (as an asymptotic series valid as $T \rightarrow 0^{+}$) with the help
of the Euler-Maclaurin formula.$^{13}$

Setting $r={n\over \sqrt{n^{2}+\sigma^{2}a^{2}}}$, one has the
asymptotic series$^{8-9}$
$$ \eqalignno{
nf(n;\sigma a) & \sim {n\sqrt{n^{2}+\sigma^{2}a^{2}}\over \sigma^{2}a^{2}}
\left[{r\over 2}{(1-r^{2})\over n}-{r^{4}\over 2}
{(1-r^{2})\over n^{2}} \right. \cr
& + {r^{3}\over 8}{(1-r^{2})(1-12r^{2}+15r^{4})\over n^{3}}\cr
& \left. + {r^{4}\over 16}
{(1-r^{2})(2-53r^{2}+168r^{4}-125r^{6})\over n^{4}}
+ ... \right] \; \; \; \; ,
&(4.16)\cr}
$$
valid as $\sigma \rightarrow \infty$, uniformly in $n$. This is derived
from the uniform asymptotic expansions of $I_{\nu}(z),K_{\nu}(z),I_{\nu}'(z)$,
and $K_{\nu}'(z)$ described in Refs. 10,11.
Denoting by ${\textstyle {L_{I}}}$ the inverse Laplace transform, one thus
has the asymptotic series
$$ \eqalignno{
L_{I}\Bigr[nf(n;\sigma a)\Bigr]& \sim
{1\over 2a^2} ne^{-{{\textstyle {n^2 T}}
\over {\textstyle {a^2}}}}
-{2T^{\textstyle
{3\over 2}}\over {3\sqrt{\pi}a^{5}}} n^3
e^{-{{\textstyle {n^2 T }}\over {\textstyle {a^2}}}}\cr
&+{T \over 8a^{4}}n
e^{-{{\textstyle {n^2 T}}\over {\textstyle {a^2}}}}
-{3T^{2}\over 4a^{6}} n^{3}
e^{-{{\textstyle {n^2 T}}\over {\textstyle {a^2}}}} \cr
&+{5\over 16}{T^3 \over a^8}
n^5 e^{-{{\textstyle {n^2 T}}\over {\textstyle {a^2}}}}
+{1\over 6\sqrt{\pi}}{T^{\textstyle
{3\over 2}}\over a^5} n
e^{-{{\textstyle {n^2 T}}\over {\textstyle {a^2}}}} \cr
&-{53\over 30}{T^{\textstyle {5\over 2}}\over
{\sqrt{\pi}a^7}} n^3
e^{-{{\textstyle {n^2 T}}\over {\textstyle {a^2}}}}
+{168\over 105}{T^{\textstyle {7\over 2}}\over
{\sqrt{\pi}a^9}} n^5
e^{-{{\textstyle {n^2 T}}\over {\textstyle {a^2}}}} \cr
&-{50\over 189}{T^{\textstyle {9\over 2}}\over
{\sqrt{\pi}a^{11}}} n^7
e^{-{{\textstyle {n^2 T}}\over {\textstyle {a^2}}}}
\; \; \; \; + ... \; \; \; \; .
&(4.17) \cr}
$$
Note that, when each term on the r.h.s. of Eq.(4.17) is
summed from $n=1$ to $N$, the resulting function of $T$
does not always converge uniformly to the sum
$\sum_{n=1}^{\infty}(\; \; \; \;)$ in a neighbourhood
$T \in (0,\delta)$ [see Eqs.(4.19)-(4.21)].  Nevertheless,
a study of the error terms shows that it is valid to take
the limit $N\rightarrow \infty$ as in Eqs.(4.19)-(4.21),
and then examine the small-$T$ behaviour of the resulting
contributions to $G(T)$.

In order to compute sums of the type
$\sum_{n=1}^{\infty}n^{\textstyle {(2m+1)}}e^{-{{\textstyle {n^2 T}}
\over {\textstyle {a^2}}}}$ where $m=0$,$1$,$2$,..., we can use the
Euler-Maclaurin formula $^{13}$
$$ \eqalignno{
{1\over 2}F(0)+F(1)+F(2)+...-\int_{0}^{\infty}F(y)dy&=
-{1\over 2}{\widetilde B}_{2}F'(0) \cr
&-{{\widetilde B}_{4}\over 4!}F'''(0) -{{\widetilde B}_{6}\over 6!}F'''''(0)
... \; \; ,
&(4.18) \cr}
$$
for the function $F(y)=ye^{-{{\textstyle {y^2 T}}\over
{\textstyle {a^2}}}}$. In Eq. (4.18), the ${\widetilde B}_{i}$ denote the
Bernoulli numbers. Thus we get
$$
\sum_{n=1}^{\infty}ne^{-{{\textstyle {n^2 T}}\over
{\textstyle {a^2}}}}={a^2 \over 2T} -{1\over 12}
-{T \over 120a^2}-{T^{2}\over 504a^4}+...
\; \; \; \; .
\eqno (4.19)
$$
The other sums arising from Eq. (4.17) are obtained by
differentiating Eq. (4.19) with respect to $T$. This yields
$$
\sum_{n=1}^{\infty}n^3 e^{-{{\textstyle {n^2 T}}\over
{\textstyle {a^2}}}}={a^4 \over 2T^{2}}+{1\over 120}
+{T \over 252a^2} +... \; \; \; \; ,
\eqno (4.20)
$$
$$
\sum_{n=1}^{\infty}n^5e^{-{{\textstyle {n^2 T}}\over
{\textstyle {a^2}}}}={a^6 \over T^{3}} -{1\over 252} +...
\; \; \; \; ,
\eqno (4.21)
$$
and so on.

Combining the resulting contribution to the asymptotic expansion of
$G^{I}(T)$ as $T \rightarrow 0^{+}$ with the other piece, arising from
the first term in Eq. (4.14), given in Ref. 9, as well as the expansion
(4.11) of $G^{F}(T)$, one finds the asymptotic expansion of the heat
kernel :
$$
G(T)\sim {a^4 \over 16}T^{-2}
+a^{3}\left({1\over 6\sqrt{\pi}}-{\sqrt{\pi}\over 8}\right)
T^{-{\textstyle {3\over 2}}}
+a\left({5\over 24\sqrt{\pi}}-{11\sqrt{\pi}\over 256}
\right)T^{-{\textstyle {1\over 2}}}
+{11\over 360} + O(\sqrt{T}) \; ,
\eqno (4.22)
$$
valid as $T \rightarrow 0^{+}$. In particular, this yields
$$
\zeta(0)={11\over 360}
\eqno (4.23)
$$
for the spin-${1\over 2}$ field with spectral boundary conditions on the
sphere.
\vskip 10cm
\centerline {\bf V. CALCULATION OF $\zeta(0)$ FOR THE
SPIN-${3\over 2}$ FIELD}
\centerline {\bf WITH SPECTRAL BOUNDARY CONDITIONS}
\vskip 1cm
In this section we sketch the corresponding calculation of
$\zeta(0)$ for a linearized spin-${3\over 2}$ field subject
to spectral boundary conditions on a 3-sphere of radius
$a$. As in the original quantum-gravity calculation of
Schleich,$^{8}$ we work only with physical degrees of
freedom by imposing a gauge condition and constraints.
This will exclude the contribution of gauge and ghost modes
which appear in the full BRST-invariant path integral.
[For a BRST-invariant approach to computing $\zeta(0)$
for a spin-${3\over 2}$ field with local boundary
conditions, see Ref.14]. The value $\zeta(0)=-
{289\over 360}$ found here for spectral boundary conditions
is again identical to that found using the natural local
boundary conditions,$^{1}$ working with the same physical
degrees of freedom inside the 3-sphere.

The spin-${3\over 2}$ field, as appearing, e.g., in $N=1$
supergravity, is described by a potential
$\Bigr(\psi_{\; \; \mu}^{A}, \;
{\widetilde \psi}_{\; \; \mu}^{A'}\Bigr)$ in the Euclidean
regime ($\mu=0,1,2,3$). In a Hamiltonian treatment,$^{15}$
the quantities $\Bigr(\psi_{\; \; i}^{A}, \;
{\widetilde \psi}_{\; \; i}^{A'}\Bigr)$ are the dynamical
variables ($i=1,2,3$), while $\Bigr(\psi_{\; \; 0}^{A}, \;
{\widetilde \psi}_{\; \; 0}^{A'}\Bigr)$ appear as Lagrange
multipliers. The gravitational field is correspondingly
described by the tetrad $e_{\; \; \; \; \; \mu}^{AA'}$.
Taking the geometry to be flat, and $x^{0}=\tau$ to be
the radial distance from the origin, while $x^{i}$
($i=1,2,3$) are coordinates on the 3-sphere, we impose the
gauge conditions
$$
e_{AA'}^{\; \; \; \; \; j}\psi_{\; \; j}^{A}=0 \; \; \; \; ,
\; \; \; \; e_{AA'}^{\; \; \; \; \; j}
{\widetilde \psi}_{\; \; j}^{A'}=0 \; \; \; \; .
\eqno (5.1)
$$
If in addition we require that the dynamical variables
$\Bigr(\psi_{\; \; i}^{A}, \;
{\widetilde \psi}_{\; \; \; i}^{A'}\Bigr)$ obey the
linearized supersymmetry constraint equations$^{15}$
on the family of 3-spheres centred on the origin, then the
expansion of $\Bigr(\psi_{\; \; i}^{A}, \;
{\widetilde \psi}_{\; \; i}^{A'}\Bigr)$ in harmonics,$^{16}$
analogous to Eqs.(1.2,3) for the spin-${1\over 2}$ field,
takes the simplified form
$$
\psi_{\; \; i}^{A}={\tau^{-{3\over 2}}\over 2\pi}
\sum_{npq}\alpha_{n}^{pq}\Bigr[m_{np}(\tau)\beta^{nqABB'}
+{\widetilde r}_{np}(\tau){\overline \mu}^{nqABB'}\Bigr]
e_{BB'i} \; \; \; \; ,
\eqno (5.2)
$$
$$
{\widetilde \psi}_{\; \; \; i}^{A'}=
{\tau^{-{3\over 2}}\over 2\pi}\sum_{npq}\alpha_{n}^{pq}
\Bigr[{\widetilde m}_{np}(\tau){\overline \beta}^{nqBA'B'}
+r_{np}(\tau)\mu^{nqBA'B'}\Bigr]e_{BB'i} \; \; \; \; .
\eqno (5.3)
$$
In the summations, $n$ runs from $0$ to $\infty$, and $p$
and $q$ now run from $1$ to $(n+1)(n+4)$. For each $n$,
$\alpha_{n}^{pq}$ is a matrix again block-diagonal in the
indices $pq$, with blocks
$\pmatrix {1&1\cr 1&-1\cr}$. Here
$$
\beta^{nqABB'}=-\rho^{nq(ABC)}n_{C}^{\; \; B'}
\; \; \; \; ,
\eqno (5.4)
$$
where $n^{AA'}=i \; {_{e}n^{AA'}}$ is the Lorentzian
normal,$^{2,7}$ and the harmonic $\rho^{nq(ABC)}$,
symmetric in its three unprimed indices, has positive
eigenvalue ${1\over 2}(n+{5\over 2})$ of the appropriate
3-dimensional Dirac operator. Similarly
$$
\mu^{nqBA'B'}=-\sigma^{nq(A'B'C')}n_{\; \; C'}^{B}
\; \; \; \; ,
\eqno (5.5)
$$
where the harmonic $\sigma^{nq(A'B'C')}$, symmetric in its
three primed indices, also has positive eigenvalue
${1\over 2}(n+{5\over 2})$. The harmonics
${\overline \mu}^{nqABB'}$ and ${\overline \beta}^{nqBA'B'}$
are given similarly in terms of harmonics
${\overline \sigma}^{nq(ABC)}$ and
${\overline \rho}^{nq(A'B'C')}$, which have negative
eigenvalues $-{1\over 2}(n+{5\over 2})$ of the 3-dimensional
operator.

Just as in the spin-${1\over 2}$ case, the simplest natural
spectral boundary conditions for spin ${3\over 2}$ are
$$
m_{np}(a)=0 \; \; \; \; , \; \; \; \; r_{np}(a)=0
\; \; \; \; , \; \; \; \; \forall n,p \; \; \; \; .
\eqno (5.6)
$$
The requirement that the physical fields be bounded near
the origin $\tau=0$, in the Hartle-Hawking path integral
for the linearized spin-${3\over 2}$ theory, implies that
$m_{np}, r_{np}, {\widetilde m}_{np}, {\widetilde r}_{np}$
vanish $\forall n,p$ at $\tau=0$. The Euclidean action
$I_{E}$ for the linearized spin-${3\over 2}$ field, working
only with the physical degrees of freedom given in
Eqs.(5.2,3), is analogous to the spin-${1\over 2}$
expression in Eqs.(2.3,4) :
$$
I_{E}=\sum_{n=0}^{\infty}\sum_{p=1}^{(n+1)(n+4)}
\Bigr[{\hat I}_{n}(m_{np},{\widetilde m}_{np})+
{\hat I}_{n}(r_{np},{\widetilde r}_{np})\Bigr]
\; \; \; \; ,
\eqno (5.7)
$$
where
$$
{\hat I}_{n}(x,{\widetilde x})=\int_{0}^{a}d\tau \;
\left[{1\over 2}\Bigr({\widetilde x}{\dot x}+
x {\dot {\widetilde x}}\Bigr)-
{(n+{5\over 2})\over \tau}{\widetilde x}x\right]
\; \; \; \; .
\eqno (5.8)
$$
The resulting Berezin integration for the Hartle-Hawking
amplitude $K_{HH}$ gives the formal expression
$$
K_{HH}=\prod_{n=0}^{\infty}\prod_{k=1}^{\infty}
{\left({{\hat E}_{nk}\over {\widetilde \mu}}\right)}
^{2(n+1)(n+4)} \; \; \; \; .
\eqno (5.9)
$$
The ${\hat E}_{nk}$ ($k=1,2,...$) are the positive eigenvalues
of the system
$$
\left({d\over d\tau}-{\hat \nu}\right)x_{nk}=
{\hat E}_{nk}{\widetilde x}_{nk} \; \; \; \; ,
\; \; \; \; \left(-{d\over d\tau}-{\hat \nu}\right)
{\widetilde x}_{nk}={\hat E}_{nk}x_{nk} \; \; \; \; ,
\eqno (5.10,11)
$$
subject to $x_{nk}(0)={\widetilde x}_{nk}(0)=x_{nk}(a)=0$,
where ${\hat \nu}={(n+{5\over 2})\over \tau}$. The solutions
are of the form
$$
x_{nk}=B_{nk}\sqrt{\tau}J_{n+2}\Bigr({\hat E}_{nk}\tau\Bigr)
\; \; \; \; ,
\eqno (5.12)
$$
$$
{\widetilde x}_{nk}=-B_{nk}\sqrt{\tau}J_{n+3}
\Bigr({\hat E}_{nk}\tau \Bigr) \; \; \; \; ,
\eqno (5.13)
$$
with the eigenvalue condition
$$
J_{n+2}\Bigr({\hat E}_{nk}a\Bigr)=0 \; \; \; \; ,
\; \; \; \; n=0,1,2,... \; \; \; \; ,
\eqno (5.14)
$$
and degeneracy $2(n+1)(n+4)$.

The formal expression (5.9) for $K_{HH}$ is then evaluated
by studying the corresponding zeta-function and heat kernel,
by analogy with Secs.II-IV. The formulae in Sec.III can be
straightforwardly modified to allow for the different
degeneracy $2(n+1)(n+4)$, different differential equations
(5.10,11), and eigenvalue condition (5.14). In particular,
Eqs.(3.14,15) should be modified by replacing the factor of
$(n+2)$ by $(n+4)$, and by changing the order $n+1$ of each
modified Bessel function $I_{n+1}$ or $K_{n+1}$ to the order
$n+2$.

For the sake of brevity, we consider only the constant part
$B_{4}=\zeta(0)$ in the expansion (3.3) of the
spin-${3\over 2}$ heat kernel $G(T)$ as $T \rightarrow
0^{+}$. The results of Eqs.(4.2)-(4.13) show that the free
part $G^{F}(T)$ of the heat kernel for spin ${3\over 2}$,
given by Eq.(4.1) with the factor $(n+1)$ replaced by
$(n+3)$, gives no contribution to $\zeta(0)$. The only
contribution to $\zeta(0)$ for spin ${3\over 2}$ arises
from the interacting part
$$
{\hat G}^{I}\left(\sigma^{2}\right)=-a^{2}
\sum_{n=2}^{\infty}(n^{2}-2)f(n;\sigma a)-a^{2}
\sum_{n=2}^{\infty}nf(n;\sigma a) \; \; \; \; ,
\eqno (5.15)
$$
where $f(n;\sigma a)$ is given by Eq.(4.15). As in
Sec.IV, these sums diverge, and the calculation should
strictly proceed by first computing the sums
$\sum_{n=1}^{N}$ for large $\sigma^{2}$, taking the
inverse Laplace transform, and finally taking the limit
$N\rightarrow \infty$. The contribution from the first
term in Eq.(5.15) is found, following Refs.8,9,12, by
using a Watson transform :
$$
-a^{2}\sum_{n=2}^{\infty}(n^{2}-2)f(n;\sigma a)=-
{a^{2}\over 4i}\int_{C'-Q}d\nu \;
(\nu^{2}-2)f(\nu;\sigma a) \cot \pi \nu \; \; \; \; .
\eqno (5.16)
$$
Here the contour of integration $C'-Q$ (compare with
Fig.1 of Ref.8) encloses all poles along the real axis,
except for those at $\nu=0,\pm 1$. Using the uniform
asymptotic expansion of $f(\nu;\sigma a)$, and
computing the contribution from the poles at
$\nu=0,\pm1$, one finds from the large-$\sigma$
behaviour of Eq.(5.16) that the first term in Eq.(5.15)
contributes $-{121\over 90}$ to $\zeta(0)$. Using the
results of Sec.IV, the second term in Eq.(5.15) is found
to contribute ${13\over 24}$ to $\zeta(0)$.

Combining these results, we find
$$
\zeta(0)=-{289\over 360}
\eqno (5.17)
$$
for the linearized spin-${3\over 2}$ field subject to
spectral boundary conditions on the 3-sphere, working
only with physical degrees of freedom in the gauge (5.1).
Just as in the spin-${1\over 2}$ case of Sec.IV, the
value of $\zeta(0)$ is equal to that found previously for
the spin-${3\over 2}$ field subject to the natural local
boundary conditions.$^{1,2}$ Specifically, the value
$\zeta(0)=-{289\over 360}$ was also found for spin
${3\over 2}$, working only with physical degrees of
freedom subject again to the gauge conditions
(5.1), with the local boundary conditions
$$
\sqrt{2} \; {_{e}n_{A}^{\; \; A'}}\psi_{\; \; i}^{A}=
\epsilon \; {\widetilde \psi}_{\; \; \; i}^{A'}
\; \; \; \; ,
\eqno (5.18)
$$
where $\epsilon=\pm 1$. These local boundary conditions
are part of a locally supersymmetric family of boundary
conditions for different spins,$^{1,2,5,14}$ which
includes the local boundary conditions (1.1) for spin
${1\over 2}$. Of course, as remarked earlier, the results
of $\zeta(0)$ calculations for gauge fields such as
spin ${3\over 2}$ will be modified by the contribution of
gauge and ghost modes in a complete BRST-invariant
calculation.
\vskip 1cm
\centerline {\bf VI. COMMENTS}
\vskip 1cm
It is striking that the same value $\zeta(0)=
{11\over 360}$ is obtained for a massless Majorana
spin-${1\over 2}$ field on a ball in Euclidean
4-space, bounded by a sphere of radius $a$, whether
local boundary conditions (1.1) or spectral boundary
conditions are imposed. The different eigenvalue
conditions, respectively,
$$
[J_{n+1}(Ea)]^{2}-[J_{n+2}(Ea)]^{2}=0 \; \;,
n=0,1,2,..., \; degeneracy \; (n+1)(n+2)
\; \; \; \; ,
\eqno (6.1)
$$
and
$$
J_{n+1}(Ea)=0 \; \; \; \; , \; \; \; \; n=0,1,2,...,
\; degeneracy \; 2(n+1)(n+2) \; \; \; \; ,
\eqno (6.2)
$$
offer no obvious explanation of this equality-
although they suggest an alternative approach through
studying the asymptotic distribution of eigenvalues.
The same holds for the value $\zeta(0)=-{289\over 360}$
found for the spin-${3\over 2}$ field (taking only physical
degrees of freedom), both for local and spectral boundary
conditions. There the eigenvalue conditions are, respectively,
$$
[J_{n+2}(Ea)]^{2}-[J_{n+3}(Ea)]^{2}=0 \; , \;
n=0,1,2,..., \; degeneracy \; (n+1)(n+4) \; \; ,
\eqno (6.3)
$$
and
$$
J_{n+2}(Ea)=0 \; \; , \; \; n=0,1,2,..., \;
degeneracy \; 2(n+1)(n+4) \; \; \; \;.
\eqno (6.4)
$$

The eigenvalue conditions for fermions, subject to
spectral boundary conditions on the 3-sphere, are
in fact similar, but not identical, to the eigenvalue
conditions for bosons, subject to local (Dirichlet)
conditions on the 3-sphere. For example, in the case
of a scalar field,$^{9,12}$ Dirichlet conditions
give
$$
J_{n+1}(Ea)=0 \; \; \; \; , \; \; \; \;
n=0,1,2,..., \; degeneracy \; (n+1)^{2} \; \; \; \; .
\eqno (6.5)
$$
For a Maxwell field, taking only physical degrees
of freedom,$^{17}$ Dirichlet (magnetic) boundary
conditions give
$$
J_{n+2}(Ea)=0 \; \; \; \; , \; \; \; \;
n=0,1,2,..., \; degeneracy \; 2(n+1)(n+3) \; \; \; \; .
\eqno (6.6)
$$
Analogous equations hold in the spin-$2$ case.$^{8}$
Because of the equality of the $\zeta(0)$ values for
fermions with both local and spectral boundary
conditions, one might then ask if there is some
connection between $\zeta(0)$ values for fermions and
bosons with adjacent spins (taking local boundary
conditions), reminiscent of supersymmetry. One would
also like to understand whether the equality of the
local and spectral values for $\zeta(0)$ is a feature
peculiar to the highly symmetrical example of a
3-sphere surrounding a region of flat 4-space, or
whether there is an extension of this result to a more
general context.
\vskip 0.3cm
\centerline {\bf ACKNOWLEDGMENTS}
\vskip 0.3cm
We are much indebted to Michael Atiyah,
Gary Gibbons, Stephen Hawking and Jorma Louko
for several enlightening discussions.
\vskip 100cm
\centerline {\bf REFERENCES}
\vskip 1cm
\item {$^{*}$}
Present address : I.N.F.N., Gruppo IV Sezione di Napoli, Italy.
\item {${ }^{1}$}
P. D. D'Eath and G. V. M. Esposito, {\it The Effect of Boundaries in One-Loop
Quantum Cosmology}, to appear in Proceedings of the Italian IX
National Conference of General Relativity and Gravitational Physics,
Capri, 25-28 September 1990, DAMTP preprint R-90/20.
\item {${ }^{2}$}
P. D. D'Eath and G. V. M. Esposito, Phys. Rev. D {\bf 43},
3234 (1991).
\item {${ }^{3}$}
P. Breitenlohner and D. Z. Freedman, Ann. Phys. {\bf 144}, 249 (1982).
\item {${ }^{4}$}
S. W. Hawking, Phys. Lett. {\bf 126 B}, 175 (1983).
\item {${ }^{5}$}
H. C. Luckock and I. G. Moss, Class. Quantum Grav. {\bf 6}, 1993 (1989).
\item {${ }^{6}$}
M. F. Atiyah, V. K. Patodi and I. M. Singer, Math. Proc. Camb. Phil. Soc.
{\bf 77}, 43 (1975).
\item {${ }^{7}$}
P. D. D'Eath and J. J. Halliwell, Phys. Rev. D {\bf 35}, 1100 (1987).
\item {${ }^{8}$}
K. Schleich, Phys. Rev. D. {\bf 32}, 1889 (1985).
\item {${ }^{9}$}
K. Stewartson and R. T. Waechter, Proc. Camb. Phil. Soc. {\bf 69}, 353
(1971).
\item {${ }^{10}$}
F. W. J. Olver, Phil. Trans. Roy. Soc. London, {\bf A 247}, 328 (1954).
\item {${ }^{11}$}
M. Abramowitz and I. A. Stegun, {\it Handbook of Mathematical Functions}
(Dover, New York, 1964).
\item {${ }^{12}$}
G. Kennedy, {\it Some Finite Temperature Quantum Field Theory Calculations
in Manifolds with Boundaries}, Ph. D. Thesis, University of Manchester, 1979.
\item {${ }^{13}$}
H. Jeffreys and B. S. Jeffreys, {\it Methods of Mathematical Physics}
(CUP, New York, 1946).
\item {${ }^{14}$}
S. Poletti, Phys. Lett. {\bf 249 B}, 249 (1990).
\item {${ }^{15}$}
P. D. D'Eath, Phys. Rev. D {\bf 29}, 2199 (1984).
\item {${ }^{16}$}
D. I. Hughes, {\it Supersymmetric Quantum Cosmology}, Ph. D. Thesis,
University of Cambridge, 1990.
\item {${ }^{17}$}
J. Louko, Phys. Rev. D {\bf 38}, 478 (1988).
\bye